\documentclass[runningheads]{llncs}
\usepackage{graphicx}
\usepackage{caption}
\usepackage{subfigure}

\authorrunning{Qi Zhang et al.}
\begin{document}
\title{LedgerGuard: Improving Blockchain Ledger Dependability \protect\footnote{This paper has been accepted by 2018 International Conference on Blockchain (ICBC)}}
%
%

\author{Qi Zhang\inst{1},
Petr Novotny\inst{1},
Salman Baset\inst{1}
Donna Dillenberger\inst{1}\\
Artem Barger\inst{2},
Yacov Manevich\inst{2}
}

%
\institute{IBM Research, Yorktown, USA \and IBM Research, Haifa, Israel
}
\maketitle              

\begin{abstract}

The rise of crypto-currencies has spawned great interest in their underlying technology, namely, Blockchain. The central component in a Blockchain is a shared distributed ledger. A ledger comprises series of blocks, which in turns contains a series of transactions. An identical copy of the ledger is stored on all nodes in a blockchain network. Maintaining ledger integrity and security is one of the crucial design aspects of any blockchain platform. Thus, there are typically built-in validation mechanisms leveraging cryptography to ensure the validity of incoming blocks before committing them into the ledger. However, a blockchain node may run over an extended period of time, during which the blocks on the disk can may become corrupted due to software or hardware failures, or due to malicious activity. This paper proposes \emph{LedgerGuard}, a tool to maintain ledger integrity by detecting corrupted blocks and recovering these blocks by synchronizing with rest of the network. The experimental implementation of \emph{LedgerGuard} is based on Hyperledger Fabric, which is a popular open source permissioned blockchain platform.

\end{abstract}

\section{INTRODUCTION}
A distributed ledger is the central component of any blockchain platform. Each peer in the Blockchain network maintains its own replica of the ledger. The ledger is an immutable append-only data structure, which contains a sequence of historical transactions grouped into blocks. The ledger is formed by chaining the blocks together with hash pointers (i.e., a subsequent block contains the hash of its previous block).

The integrity of the ledger is essential for correct functioning of the peer. With corrupted ledger, the peer is not able to generate valid transactions when the smart contract needs to retrieve historical transactions from the ledger. Moreover, when historical transactions recorded in the ledger are requested by external tools such as analytical or auditing applications, the peer first verifies the integrity and validity of the relevant blocks before extracting the transactions. Any corruption of blocks discovered by these operations leads to significant degradation of the peer functionality. Typically, the peer will cease to function till the correct ledger is available. Furthermore, the applications accessing the corrupted data may become significantly impaired as well.
 
The peer protects its ledger from introducing corrupted data. When a new block is received the peer validates the integrity of the block before appending the block to the ledger. However, the peer lacks the capability of detecting and recovering the corrupted blocks existing in the ledger during its runtime.

A corruption of the ledger may have one of several different causes. Various types of ledger corruptions have been observed on the public Blockchain platforms, such as Bitcoin \cite{nakamoto2008bitcoin} and Ethereum \cite{buterin2014next}. For instance, on Ethereum platform users reported corrupted data files due to false positives of antivirus software \cite{antivirus-corrupt}. Ledger corruptions were also reported by Bitcoin users due to block checksum mismatch \cite{bitcoin-block} . In private Blockchains such as Hyperledger Fabric \cite{androulaki2018hyperledger} or R3 Corda \cite{brown2016introducing}, it is critical to maintain the nodes hosting peers highly secure. However, when a peer is hosted in a less secure environment, an external attacker or malicious user can hack into the peer node and modify the content of the ledger files. Moreover, since the ledger files are typically stored on a storage medium such as magnetic disks or SSDs, a hardware failure \cite{gray2007empirical} \cite{pinheiro2007failure} \cite{meza2015large} may also cause a corruption of the ledger files. 

In this paper, we introduce \emph{LedgerGuard}, a mechanism that enables the peer to maintain the integrity of its ledger. \emph{LedgerGuard} enforces the integrity of the ledger with the following two techniques. First, it validates the content of each block and the hash links between blocks. Second, if corrupted block is identified, \emph{LedgerGuard} recovers the block and corrects the affected part of ledger without the need for rebuilding the whole ledger. \emph{LedgerGuard} is designed in a highly configurable manner. It can be used as a tool (e.g., by an operator) to validate and correct online or offline ledger. It can also be used as a service of the peer node, to continuously monitor and correct the ledger and thus increase the resiliency and availability of the peer.

\section{BACKGROUND} \label{Background}
In this section, we briefly describe the ledger design of Hyperledger Fabric. Although the ledger design varies among different Blockchain platforms, they follow the same principles. Due to the space limitation, we do not describe the details of Hyperledger Fabric design in the paper, but we recommend readers to refer to \cite{cachin2016architecture} \cite{vukolic2017rethinking} \cite{sousa2017byzantine} for more information. 
\begin{figure}
  \centering \includegraphics[width=8cm,height=4cm]{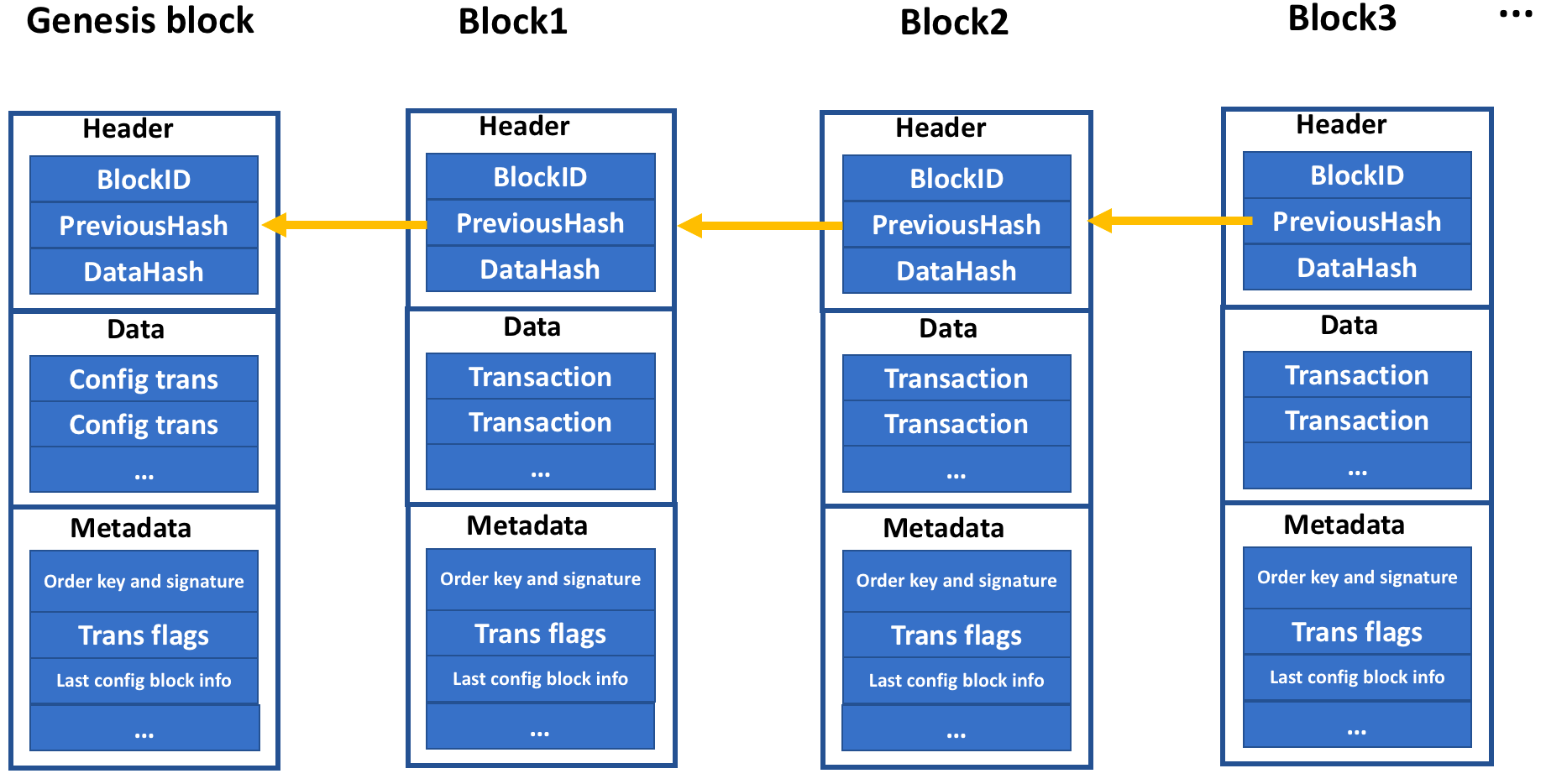}
  \caption{Hyperledger Fabric Blockchain ledger}
  \label{ledger_blocks}
\end{figure}

In Hyperledger Fabric, a transaction is submitted by the client and endorsed by multiple peers. If being successfully endorsed, the transaction with its endorsements will be further sent to the orderer, who collects transactions from multiple clients and organizes them into blocks. After that, the orderer delivers the block to the peers, who finally validates the block and commits it into ledger.

Figure \ref{ledger_blocks} shows the design of the Blockchain ledger in Hyperledger Fabric, which consists of a chain of blocks that are connected by hash pointers. Normally, a block will never be changed after being committed into the ledger. The first block is called the genesis block, which contains configuration information of this Blockchain network. Blockchain configuration can be changed over time, for example, when a new peer joins or an existing peer leaves. The new configuration transactions will be recorded in the other blocks. 

Each block has three sections: block header, block data, and block metadata. The block header section includes the sequence number of this block, the hash value of the previous block, and the hash value of the data section in the current block. The block data section contains a series of transactions with some additional information such as the read/write sets and the endorsers' signatures. For the medadata section, it incorporates the certificate, public key and the signature from the orderer. When creating the block, the orderer signs the block header and stores the signature into the metadata section. Depending on the architecture, a block can be signed by a single or multiple orderers. The metadata section also contains information such as the flags of the validity of each transaction in the block.

\section{LEDGER CORRECTION} \label{LedgerGuard}
In this section, we describe \emph{LedgerGuard}, which improves the Blockchain ledger dependability by providing a runtime self-correction mechanism for ledger.

\vspace*{1mm}
\noindent
\textbf{Approach overview. }
In order to minimize the negative impact brought by the corrupted ledger, we introduce a runtime self-correction mechanism, \emph{LedgerGuard}, for the Blockchain ledger. \emph{LedgerGuard} runs as a service on each peer, checks the integrity of the ledger on the peer, and recovers the corrupted block if there is any. We provide several options for the users to activate \emph{LedgerGuard}. First, it can be setup as a periodically running process in the peer, which is initialized when the peer starts and runs after every period of time. Second, in order to not affect the peer performance, \emph{LedgerGuard} can be triggered by a resource monitor in the peer when the hardware resource utilization, such as CPU, is under a pre-configured threshold. Third, \emph{LedgerGuard} can be provided as a peer built-in tool and explicitly activated by the user when he or she wants to know the integrity of the ledger. 

\noindent
\textbf{Ledger corruption detection} 
As shown in Figure \ref{ledger_blocks}, the blocks in the ledger are connected by the hash pointers. \emph{LedgerGuard} validates the ledger integrity from two aspects: (1) each single block in the ledger is not corrupted, and (2) the hash pointers between the blocks are valid. In Hyperledger Fabric, a block is created by the ordering service, which signs the block header and stores the signature in the block metadata. Therefore, \emph{LedgerGuard} uses the certificate of the ordering service to validate the correctness of each block header. Since the block header contains the hash value of the block data section and the signature is collected from the block metadata section, a successfully verified signature indicates the block has not been tampered with. We assume the root Certificate Authority in the Hyperledger Fabric Blockchain platform is trusted, thus \emph{LedgerGuard} can get a valid ordering service certificate to validate the blocks. To validate the correctness of the hash pointer, \emph{LedgerGuard} calculates the hash value of the current block (e.g. Hash(block X)), and compares this hash value with the value of "PreviousHash" in the header of block X+1. The hash pointer is integrated if these two value matches. Otherwise, at least one of the two blocks are corrupted. 

\noindent
\textbf{Corrupted ledger recovery} 
Once a corrupted block is detected by a specific peer (e.g. peer 1), \emph{LedgerGuard} on this peer will send a request to the other peers (e.g. peer 2) in the same Blockchain network, and ask for the block with the same ID as the corrupted block. Since all the peers have the same copy of the ledger, after peer 1 obtains the block from peer 2, it use the approach described in the previous subsection to validate the correctness of this newly received block. If this block is invalid, peer 1 will keep asking the other peers for the same block. Otherwise, peer 1 uses this newly received block to recover the corrupted ledger.
\begin{figure}
  \centering \includegraphics[width=10cm,height=4cm]{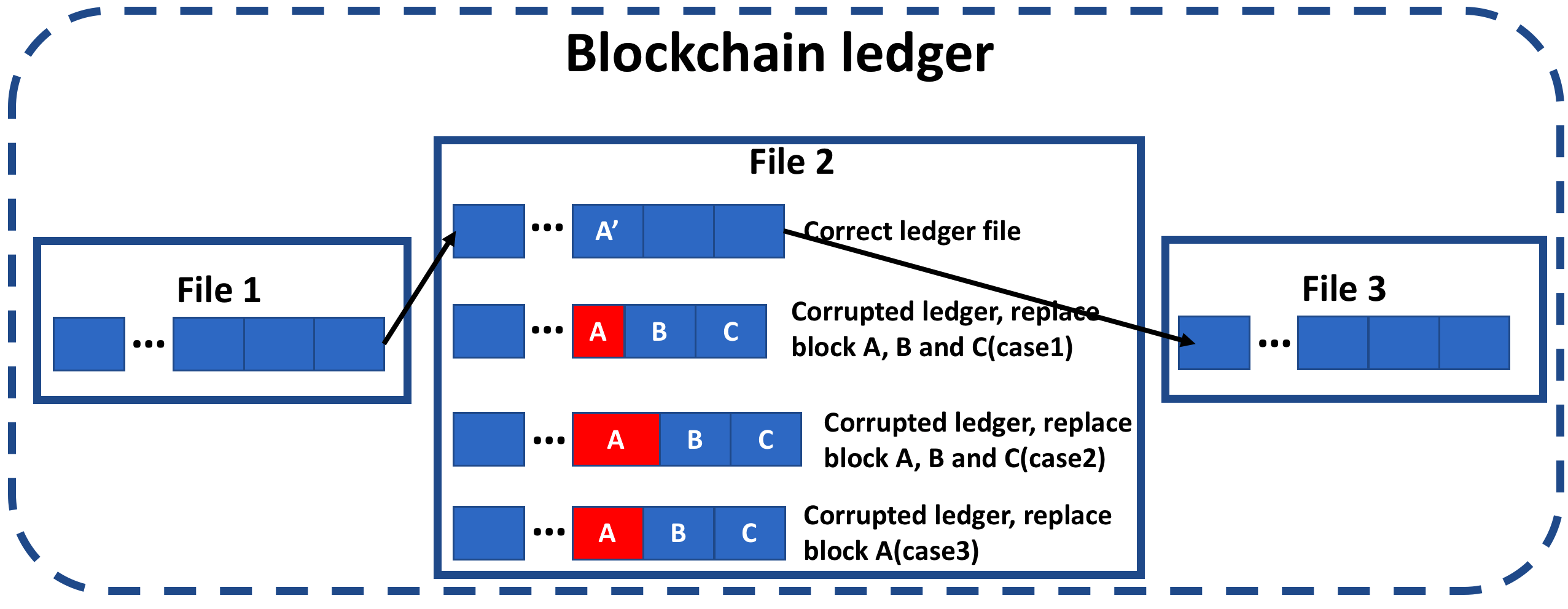}
  \caption{Blockchain ledger stored in files}
  \label{ledger_files}
\end{figure}
Sometimes multiple blocks need to be retrieved to fix the ledger even though only one block is corrupted. In Hyperledger Fabric, a ledger consists of one or multiple fix sized files, and each file contains a continuous series of blocks. A corrupted block can be either larger or smaller than the original block, thus simply replacing the corrupted block with a correct block still breaks the integrity of the ledger. As an example depicted in Figure \ref{ledger_files}, block A in file 2 is detected as a corrupted block. Block A' is a correct block retrieved from another peer. If the size of block A is not equal to that of block A', simply replacing block A with block A' will either overwrite part of block B or leave a gap between block A and block B. In order to solve this problem, \emph{LedgerGuard} first checks whether the size of block A has changed. If it is, as shown in Figure 2, the process will replace all the blocks in file 2 that are subsequent to block A (case1 and case2). Otherwise, only block A needs to be overwritten (case3).

\noindent
\textbf{Optimization}
As an in progress work, we are exploring optimizations for \emph{LedgerGuard}. For example, since hash value calculation and signature verification are CPU intensive, \emph{LedgerGuard} can use file level verification to decrease its CPU resource consumption. Concretely, when \emph{LedgerGuard} validates the ledger for the first time, it temporarily keeps the validated blocks in memory until all the blocks in a file have been validated. If all the blocks as well as the hash pointers are correct, \emph{LedgerGuard} calculates the hash value of the whole file. The hash values of the files will be kept by system administrators in a separate secure storage. Therefore, when the same portion of the ledger needs to be checked for a second time, only the hashes of the files need to be calculated and compared. Since a file usually contains many blocks, this will largely reduce the amount of hash value calculations. The linkage between the two files can be validated by checking hash pointer between the last block of the previous file and the first block of the next file.

\section{Evaluation} \label{Evaluation}
In this section, we evaluate the effectiveness of \emph{LedgerGuard} on a 4-core VMWare virtual machines, with Intel(R) Xeon(R) CPU E5-2698 2.20GHz with 4GB of RAM. The ledgers used in this section is generated by a tool, which closely simulates the blocks generation on a real Hyperledger Fabric Blockchain network. The tool first loads peer and orderer private keys and certificates, then crafts transaction proposals and endorsements signed by the peer private key, and finally batches the resulting transactions into blocks signed with the orderer private key.  A Hyperledger Fabric Blockchain network used in this section is setup with 4 peers, and each peer loads the generated ledger.

\begin{figure}
  \centering \includegraphics[width=8cm,height=4cm]{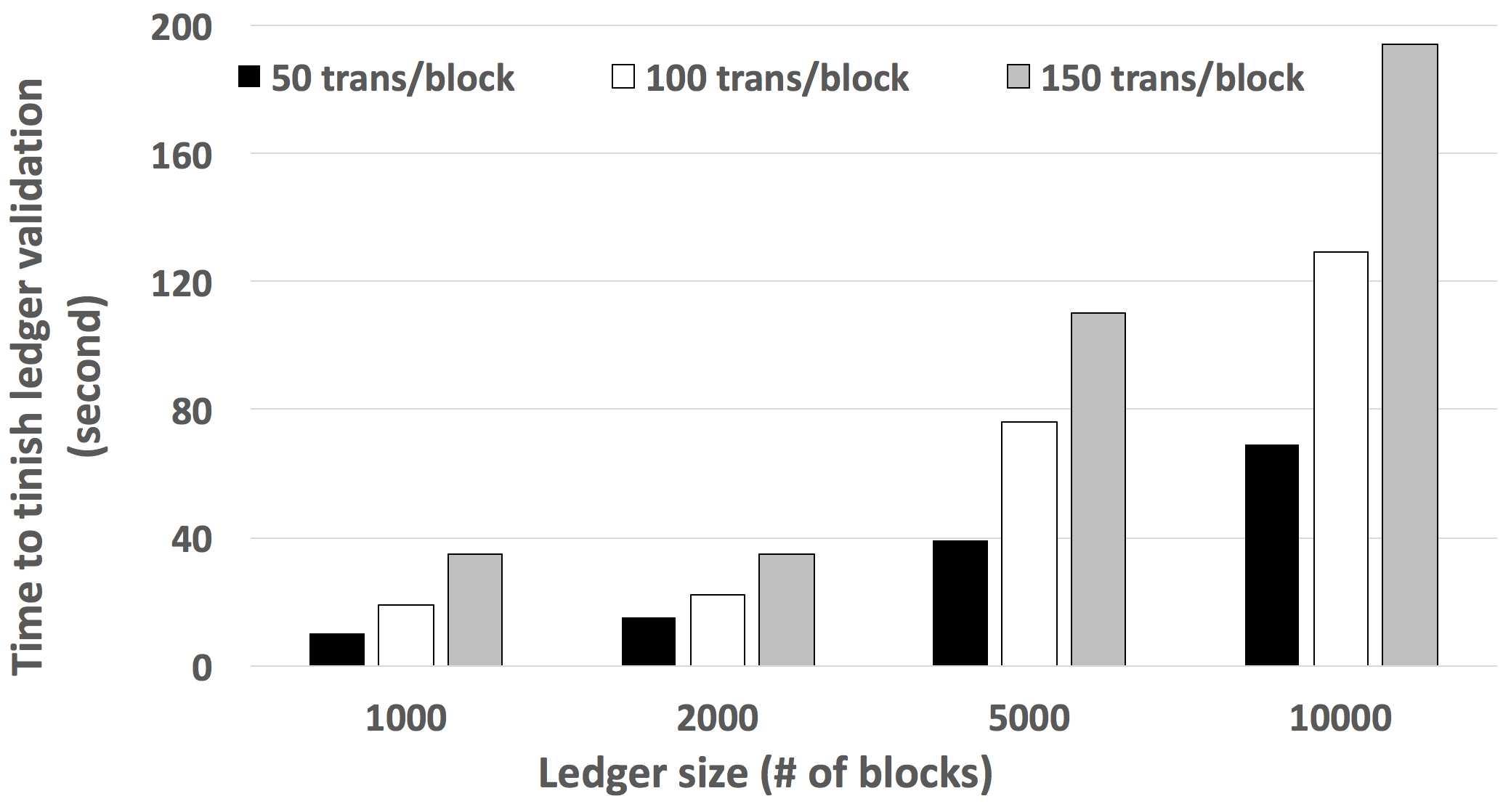}
  \caption{\emph{LedgerGuard} ledger validation time}
  \label{ledger_validation_time}
  \vspace{-0.5cm}
\end{figure}

Figure \ref{ledger_validation_time} shows how much time \emph{LedgerGuard} takes to validate all the blocks in the ledger. The ledger size of 1000 blocks, 2000 blocks, 5000 blocks, and 10000 blocks are used. For each ledger size, we vary the block size from 50 transactions per block to 150 transactions per block, and each transaction is 3KB. We observe that first, with different ledger sizes but the same block size, the larger the ledger is, the longer it takes to finish validation. For example, with 50 transactions per block, it takes 69 seconds to finish the validation of the ledger with 10000 blocks, while it takes 69 seconds when there are 5000 blocks in the ledger. This is because \emph{LedgerGuard} sequentially scans through each block in the ledger, the more blocks the ledger contains, the longer time \emph{LedgerGuard} takes to finish validation. Second, with the same ledger size but different block sizes, the larger the block size is, the longer it takes for validation. Taking the ledger with 5000 blocks as an example, it takes 39 seconds to finish validation when each block contains 50 transactions, while that number increases to 110 seconds when each block includes 150 transactions. Our measurement shows that the block hashing time does not vary much when the block size increases from 50 transactions to 150 transactions, and also the order signature verification time is independent of the block size. Therefore, the difference in the ledger validation time is mostly because the larger the block size is, the more time is spent on I/O to read the blocks.

\begin{figure}
  \centering
  \subfigure[50 trans/block]{
    \label{fig:subfig:a} 
    \includegraphics[width=1.5in, height=1.2in]{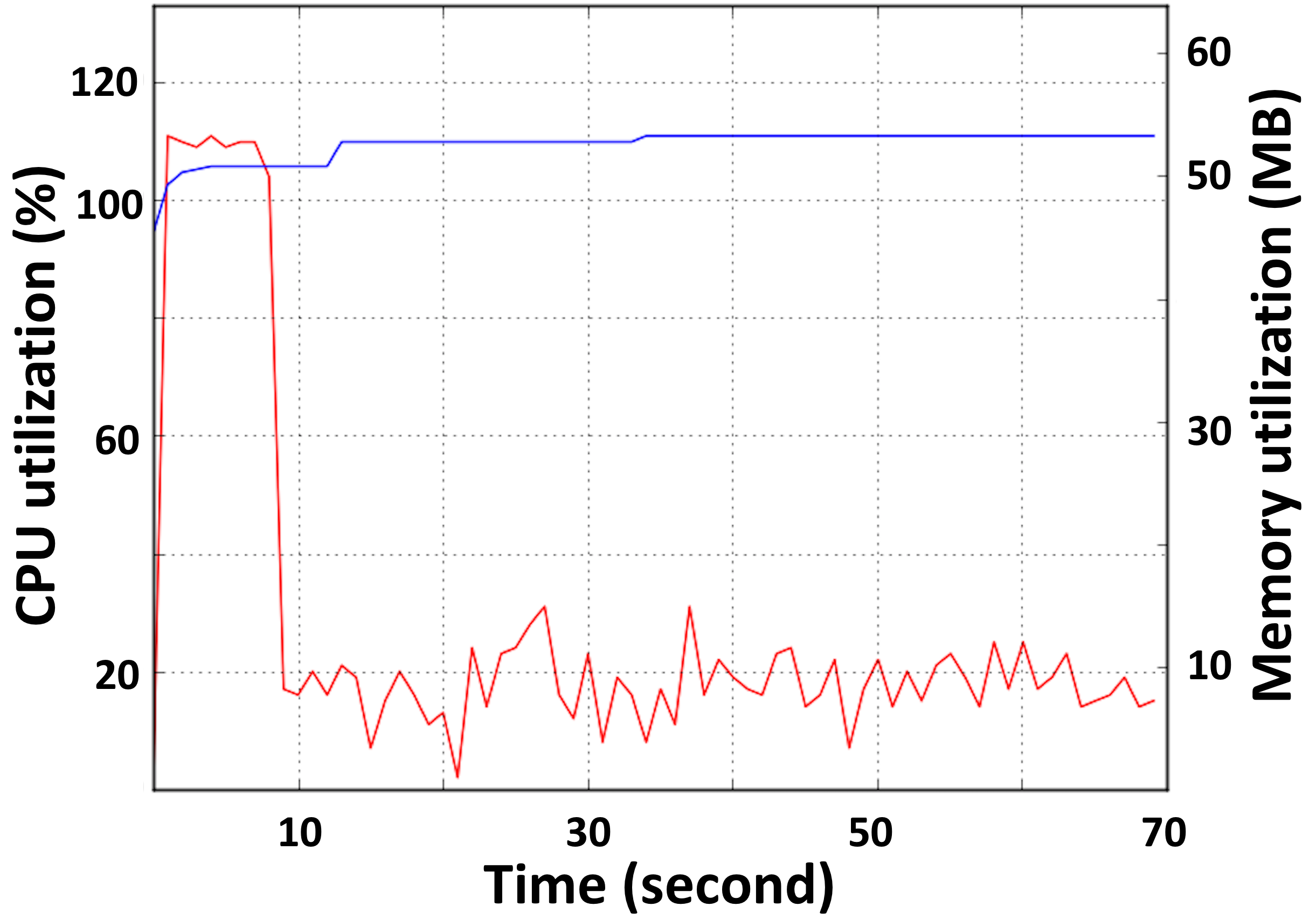}}
  \subfigure[100 trans/block]{
    \label{fig:subfig:b} 
    \includegraphics[width=1.5in, height=1.2in]{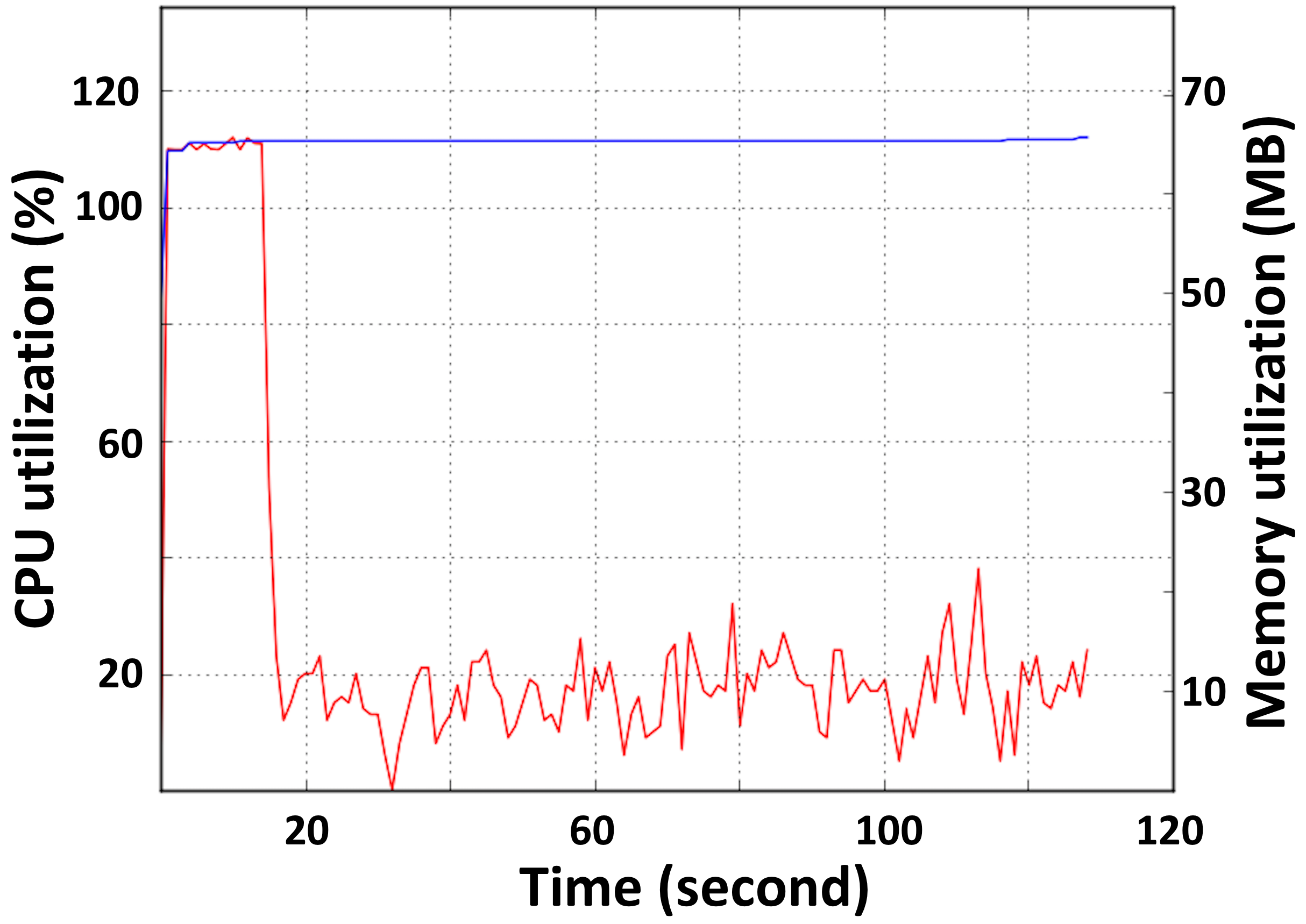}}
  \subfigure[150 trans/block]{
    \label{fig:subfig:b} 
    \includegraphics[width=1.5in, height=1.2in]{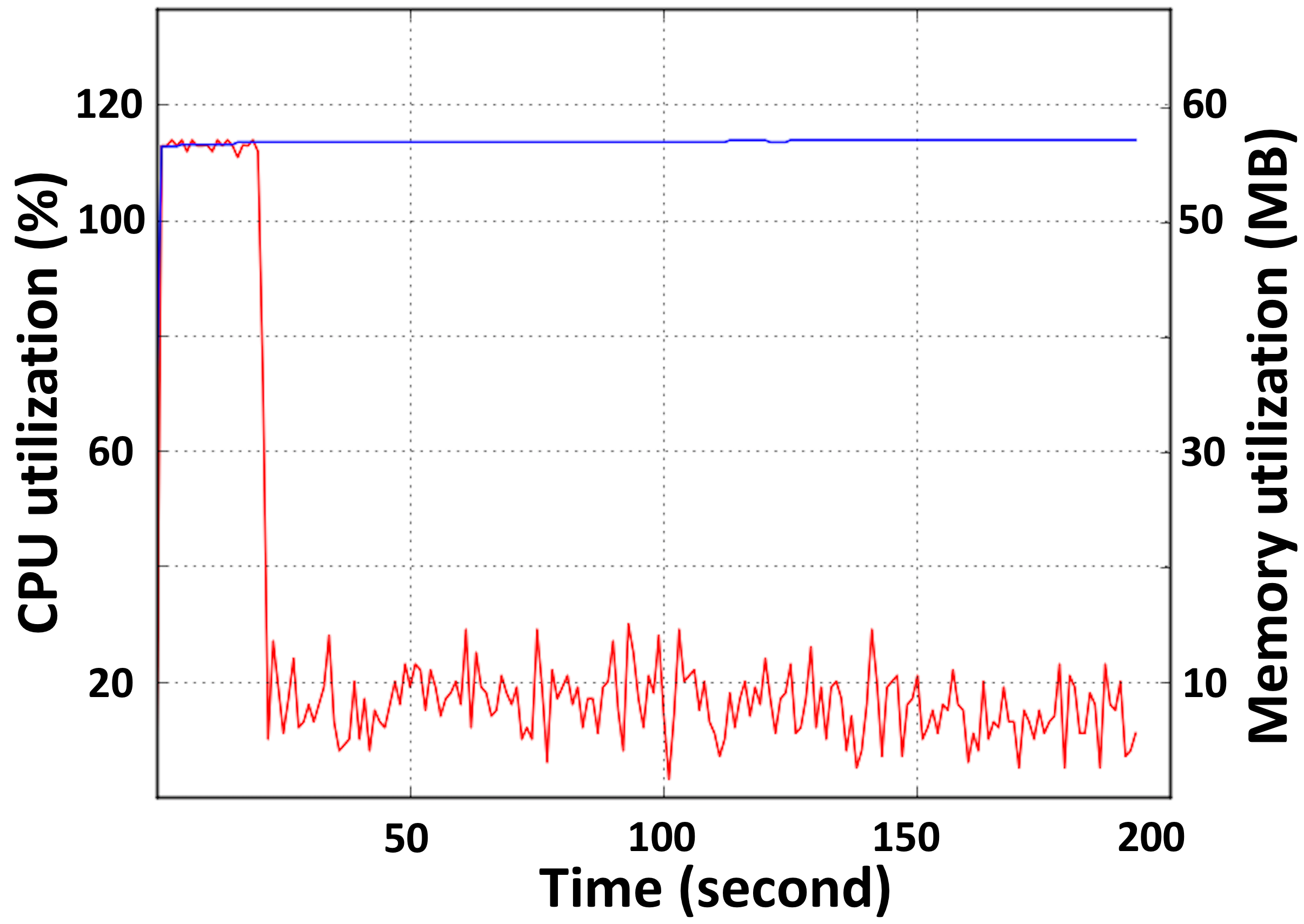}}
  \caption{\emph{LedgerGuard} CPU and memory utilization(ledger size is 10000 blocks)}
  \label{ledgerguard_resource} 
  \vspace{-0.5cm}
\end{figure}

We also measure how much CPU and memory does \emph{LedgerGaurd} consume during ledger validation, and the results are depicted in Figure \ref{ledgerguard_resource}. It shows that, no matter for what block size, \emph{LedgerGuard} uses about 60MB memory, and the CPU utilization of \emph{LedgerGuard} starts with around 110\%, then drops to around 20\% and stays stable. The reason for the initial CPU utilization spike is \emph{LedgerGuard} needs to do some initialization work such as opening the ledger, reading configuration of the blockchain network, and initializing the MSP (Membership Service Provider) manager. After that, the \emph{LedgerGuard} works as a single process to scan through the ledger and validate each block. Since calculating the block hash value and verifying the signature are both CPU intensive, the \emph{LedgerGuard} occupies the whole CPU core, which leads to around 20\% CPU utilization in a 4 core machine.

Moreover, we measure the speed of recovering the ledger. It shows that with the size of 150 transactions per block, a peer node can fetch the block from the other peer, validate and commit it in a speed of 8.5 blocks per second. As part of our on going research, we are working on creating a ledger with different distribution of corrupted blocks, and measure the effectiveness of \emph{LedgerGuard} to recover the corrupted ledger.

\section{Related Work} \label{RelatedWork}
As blockchain technologies gain popularity, issues about the Blockchain platform reliability and security have been observed. Some Ethereum users reported that the Blockchain ledger on his or her machine has been corrupted due to a false positive of antivirus software~\cite{antivirus-corrupt}. This was confirmed by another user who has seen report saying that an antivirus software corrupted an Ethereum Blockchain by deleting some file from the ledger. The suggested solution was to delete the ledger data, and restart the client to re-download the whole ledger. Error of "block checksum mismatch'' has also been observed by users of Bitcoin~\cite{nakamoto2008bitcoin}, Litecoin~\cite{litecoin}, and Dogecoin~\cite{dogecoin} when Btrfs \cite{rodeh2013btrfs} is used. The reason was due to single-bit errors when reading from disk, and the proposed solution was to change the filesystem to EXT4 \cite{cao2007ext4} and re-downloading the whole ledger~\cite{bitcoin-block}. Moreover, since smart contracts are programs that could move large value assets on the Blockchain, they always become the victims of attackers who want to steal the assets. The DAO attack \cite{dao} showed that a program built on the Ethereum Blockchain platform was breached in a case that results in \$50 million worth of Ether being stolen. Researchers and practitioners are making great efforts to improve the reliability and security of the Blockchain platform. Nicola \cite{atzei2017survey} did a survey of the attacks on Ethereum smart contacts by exploiting a series of attacks and providing a taxonomy of programming pitfalls which can lead to such vulnerabilities. Zcash \cite{hopwood2016zcash} was invented by creating transactions that reveal neither the payment's origin, destination, nor the amount. This approach prevents leakage of the users' spending habits by Blockchain mining \cite{chainalysis}.

\section{Conclusions and Future work} \label{Conclusions}
Blockchain ledger can be corrupted due to many reasons, and ensuring the integrity of the ledger is critical to the functionality and the performance of the Blockchain platform. In this paper, we propose \emph{LedgerGuard} - a mechanism to keep track of ledger integrity by detecting corrupted blocks and recover the ledger by synchronizing it with rest of the network, implement a preliminary prototype, and evaluate its effectiveness and overhead. As the on-going and future work, we are extending and improving the \emph{LedgerGuard} from multiple aspects. For example, we are exploring algorithms to enable the \emph{LedgerGuard} to smartly select the other peers based on the network connection quality when it tries to fetch a block, which will further improve the performance of recovering the ledger. Furthermore, we are investigating more alternative approaches to detect corrupted blocks other than sequential scan.

\bibliographystyle{splncs04}
\bibliography{Main} 

\end{document}